\documentclass[twocolumn,showpacs,preprintnumbers,amsmath,amssymb,aps]{revtex4}
\usepackage{graphicx}
\usepackage{dcolumn}
\usepackage{bm}

\begin{document}

\title{Narrowing of resonances in electromagnetically induced transparency and absorption using a Laguerre-Gaussian control beam}
 \author{Sapam Ranjita Chanu}
 \author{Vasant Natarajan}
 \email{vasant@physics.iisc.ernet.in}
 \homepage{www.physics.iisc.ernet.in/~vasant}
 \affiliation{Department of Physics, Indian Institute of
 Science, Bangalore 560\,012, India}

\begin{abstract}
We study the phenomenon of electromagnetically induced transparency and absorption (EITA) using a control laser with a Laguerre-Gaussian (LG) profile instead of the usual Gaussian profile, and observe resonances with width up to 20 times below the natural linewidth. Aligning the probe beam to the central hole in the doughnut-shaped LG control beam allows simultaneously a strong control intensity required for high signal-to-noise ratio and a low intensity in the probe region required to get narrow resonances. Experiments with a second-order LG beam show that transit time and orbital angular momentum do not play a significant role. This explanation is borne out by a density-matrix analysis with a radially varying control Rabi frequency. We observe these resonances using degenerate two-level transitions in the $D_2$ line of Rb in a room temperature vapor cell. We show that the linewidth is reduced by about a factor of 4 compared to the use of a Gaussian control beam, which should prove advantageous in all applications of EITA and other kinds of pump-probe spectroscopy.
\end{abstract}

\pacs{42.50.Gy,32.80.Qk,32.80.Xx}


\maketitle

\section{Introduction}
The Laguerre-Gaussian (LG) beam, also known as a doughnut mode because the phase singularity of the electric field leads to a hole in the center, has found important applications in several areas of physics. For example, the orbital angular momentum associated with the photons in the LG beam \cite{ABP03} has been used to rotate optically trapped microscopic particles \cite{PMA01} or create quantized vortices in a Bose-Einstein condensate \cite{ARC06}. LG beams have been used to create a waveguide for atoms \cite{TFH99,KAA00}, and to reduce the linewidth of Hanle resonances \cite{ARP10}. Here, we show that the LG beam can be used to create narrow resonances at the line center of an optical transition {\em with an unprecedented reduction in linewidth of 20 times below the natural linewidth}. We use the LG beam as a control beam in the well-known phenomenon of electromagnetically induced transparency and absorption (EITA) \cite{HAR97,FIM05,CSB11}, a phenomenon in which the strong control beam is used to modify the properties of a medium for a weak probe beam. We show that the LG control beam causes a significant narrowing compared to the usual Gaussian control beam. EIT has wide-ranging applications such as lasing without inversion \cite{AGA91}, nonlinear optics \cite{HFI90}, slowing of light \cite{HHD99} (for use in quantum-information processing), and white-light cavities \cite{WUX08} (for use in gravitational wave detection). EIA has been proposed for applications such as sub to superluminal switching of the probe pulse \cite{CJG07}, and giant enhancement of the Kerr nonlinearity \cite{NGL05}. In many of these applications, it is the anomalous dispersion near the EITA resonance that is responsible for the effects. Thus, the narrower the EITA resonance, the more pronounced these effects. Narrow EITA resonances also play an important role in high-resolution spectroscopy \cite{RAN02,DAN05} and tight locking of lasers to optical transitions.

The underlying mechanism for EITA is the shift of the energy levels of the atom away from line center through the AC Stark effect (creation of {\em dressed states} \cite{COR77} and an Autler-Townes doublet). The shift is equal to the Rabi frequency of the control laser, therefore the EITA resonance is subnatural when the power in the control laser is sufficiently small---e.g.\ a resonance width of $0.25 \, \Gamma$ has been seen with a control Rabi frequency of $0.3 \, \Gamma$ \cite{LIX95}. However, a {\em small} control power implies a correspondingly low signal-to-noise ratio (SNR) in the probe spectrum. As a consequence, in the above observation of the $0.25 \, \Gamma$ linewidth, the EIT dip is barely visible above the noise. The LG control beam is ideally suited to overcome these conflicting requirements; by aligning the probe beam to the center of the LG control beam, the control power can be increased for high SNR while maintaining a negligibly small power in the region of the probe required for the narrow resonance. Physically, the doughnut structure of the control beam leads to a spatial variation of the control Rabi frequency, with a low frequency in the region where the absorption of the probe beam is significant. This explanation is borne out by a detailed density-matrix analysis using a radially varying control Rabi frequency, which shows that the LG profile gives a smaller linewidth compared to the Gaussian profile. The spatial {\em separation} of the intense part of the two beams also allows the probe beam to be detected with minimal contamination from the LG control, which is often a problem with a control beam that has a Gaussian profile. Thus, the use of an LG beam should prove advantageous in several other kinds of pump-probe spectroscopy.

EITA is usually studied in three-level or multilevel systems, with the control and probe lasers on separate dipole-allowed transitions. We have recently shown the this can be observed in a ``two-level system'' \cite{CSB11}, i.e.\ one in which both lasers are on the same transition and no third level is involved. Of course, each level is composed of multiple magnetic sublevels (making it a ``multilevel system''), but this is true of all EITA experiments, even those that are called three-level systems for example. Indeed, one of the transitions we use is from $F=1$ to $F'=0$, which is the simplest transition in terms of the number of sublevels involved. In our earlier work using a Gaussian control beam \cite{CSB11}, we demonstrated the subnatural features on the $^{87}$Rb $D_2$ line, and showed that the narrow resonance appears as enhanced absorption (EIA) for $F=2 \leftrightarrow F'$ transitions and enhanced transparency (EIT) for $F=1 \leftrightarrow F'$ transitions. Here also we observe the same difference but the features are much narrower. For the $F=2 \leftrightarrow 1$ transition, the resonance has a full width that is 20 times below the natural linewidth, which as far as we know is the narrowest feature ever observed for a probe laser on a dipole-allowed transition. (The narrowest feature reported previously is $\Gamma/7$ also from our laboratory \cite{IKN08}). Considering that the LG beam can be produced quite easily by using a printout of a computer-generated hologram, the use of LG control beams should vastly increase the applications of EITA.

The LG beam is known to carry orbital angular momentum, therefore one might expect that the linewidth reduction is due to the torque on the atoms resulting in increased transit time within the control beam. To understand whether the transit time limits the linewidth, we have studied the EITA phenomenon with an expanded Gaussian control beam (thereby increasing the time that atoms spend in the beam), and with a second-order LG control beam (thereby increasing the torque on the atoms by a factor of two). We find no reduction in linewidth, which shows that {\em transit time does not play a significant role}.

\section{Experimental details}
\begin{figure}
\centering{\resizebox{0.90\columnwidth}{!}{\includegraphics{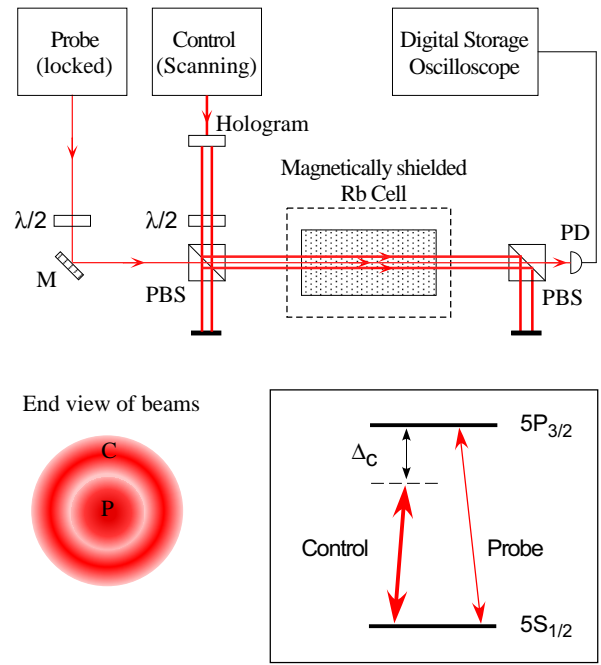}}}
\caption{(Color online) Schematic of the experiment. Figure key: $\lambda/2$ -- halfwave retardation plate, M -- mirror, PBS -- polarizing beamsplitter cube, PD -- photodiode. The inset shows the relevant energy levels of Rb used in the experiment.}
 \label{schema}
\end{figure}

The energy levels of the $5S_{1/2} \leftrightarrow 5P_{3/2}$ transition ($D_2$ line) in $^{87}$Rb used for the experiments are shown in the inset of Fig.\ \ref{schema}. The upper state has a natural linewidth $\Gamma = 2\pi \times 6$~MHz. There are two hyperfine levels ($F=1,2$) in the ground state and four hyperfine levels ($F'=0,1,2,3$) in the excited state, thereby allowing two sets of transitions: $F=1 \rightarrow 0$, 1, or 2 (lower ground-level transitions), and $F=2 \rightarrow 1$, 2, or 3 (upper ground-level transitions). Both the control and the probe lasers are on the same two-level transition.

The set of Laguerre-Gaussian modes (LG$^l_p$) forms a basis set to describe paraxial laser beams \cite{SIE86}. The indices $l$ and $p$ are the winding number (the number of times the phase completes $2\pi$ on a closed loop around the axis of propagation) and the number of radial nodes for radius $\rho>0$, respectively. We use the LG$^{1}_{0}$ mode, which has an electric field amplitude at the beam waist (in cylindrical coordinates) given by,
\begin{equation}
{\rm{LG}}^{1}_{0}\left(\rho , \phi \right)=E_0
\frac{\rho}{w_{0}} \exp \left( - \frac{\rho ^2}{w_{0}^{2}}
\right)\exp \left( i \phi \right) \, ,
 \label{lg}
\end{equation}
so that the peak-to-peak diameter is $\sqrt{2}w_{0}$.

The experimental schematic, shown in Fig.\ \ref{schema}, is similar to that in our earlier work \cite{CSB11}, except that the control beam has an LG profile. It is generated by diffracting the Gaussian beam output of a grating-stabilized diode laser \cite{BRW01} through a computer-generated transmission hologram \cite{BVS90a,HMS92}. The laser frequency is scanned by electronically varying the angle of the grating using a piezoelectric transducer. The control beam power in the experimental cell is $0.3$~mW and its $w_0$ size is 1.4~mm. The probe beam is derived from a similar but {\em independent} diode laser system. Part of the beam is sent through an acousto-optic modulator (AOM) into a saturated-absorption spectrometer. The frequency of the AOM is modulated at 30~kHz to generate the error signal for locking the laser to a peak. Frequency modulation using the AOM rather than by direct modulation of the diode current is important in keeping the laser linewidth as small as possible, and observing the narrowest features. The two beams are linearly polarized in orthogonal directions so that they can be mixed and separated using polarizing beam-splitter cubes, and only the probe beam can be detected.

The probe beam is expanded to a large size and then apertured to a diameter of about 1~mm. This ensures that the intensity is roughly constant over this size. The total power in the beam entering the cell is 6~$\mu$W. It is aligned to the central hole of the LG control beam, as shown in the figure. The two beams {\em co-propagate} through a room temperature vapor cell of size 25-mm diameter $\times$ 50-mm long. The cell contains pure Rb (both isotopes in their natural abundances and with no buffer gas), and has a multilayer magnetic shield around it. The linear absorption of the probe beam through the cell is about 2\%.

\section{Results and Discussion}
\begin{figure}
\centerline{\resizebox{0.95\columnwidth}{!}{\includegraphics{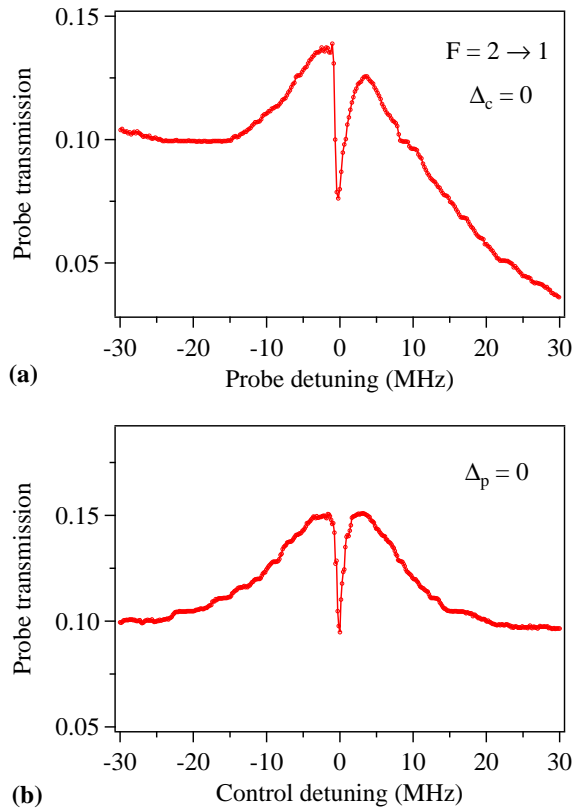}}}
\caption{(Color online) Advantage of scanning the control laser. The spectrum obtained by scanning the probe laser with a locked control [shown in (a)] appears on a Doppler background, whereas the spectrum obtained by scanning the control laser with a locked probe [shown in (b)] appears on a flat Doppler-free background.}
 \label{pvsc}
\end{figure}

Before turning to the results, we first demonstrate the advantages of scanning the control laser while keeping the probe laser locked. Any spectrum is obtained by measuring the response of a system while scanning the input frequency. Most EITA spectra are obtained by looking at probe transmission while scanning its frequency, {\em while the control beam is locked} \cite{LIX95}. Instead, we use a technique where the probe laser is locked on resonance while the detuning of the control laser is scanned \cite{DAN05}. This is a technique that we have developed to overcome the first-order Doppler effect so that the spectrum appears on a flat background. The underlying reason is that the locked probe beam addresses only the zero-velocity atoms, and its absorption remains {\em Doppler free} until it is modified by the control beam. This advantage is seen clearly in Fig.\ \ref{pvsc}. The probe-transmission spectrum for the $F=2 \rightarrow 1$ transition is obtained either with $\Delta_c = 0$ (zero control detuning) and $\Delta_p$ scanning [in (a)], or with $\Delta_p = 0$ (zero probe detuning) and $\Delta_c$ scanning [in (b)]. In both cases, the linewidth is similarly subnatural (about 1 MHz or $\Gamma/6$), but the flat background obtained with control scanning makes the {\em lineshape} much better. Therefore, all the results presented below were obtained by scanning the control laser, as has been the case with most of our recent work \cite{CPN12}.

\begin{figure}
\centerline{\resizebox{0.95\columnwidth}{!}{\includegraphics{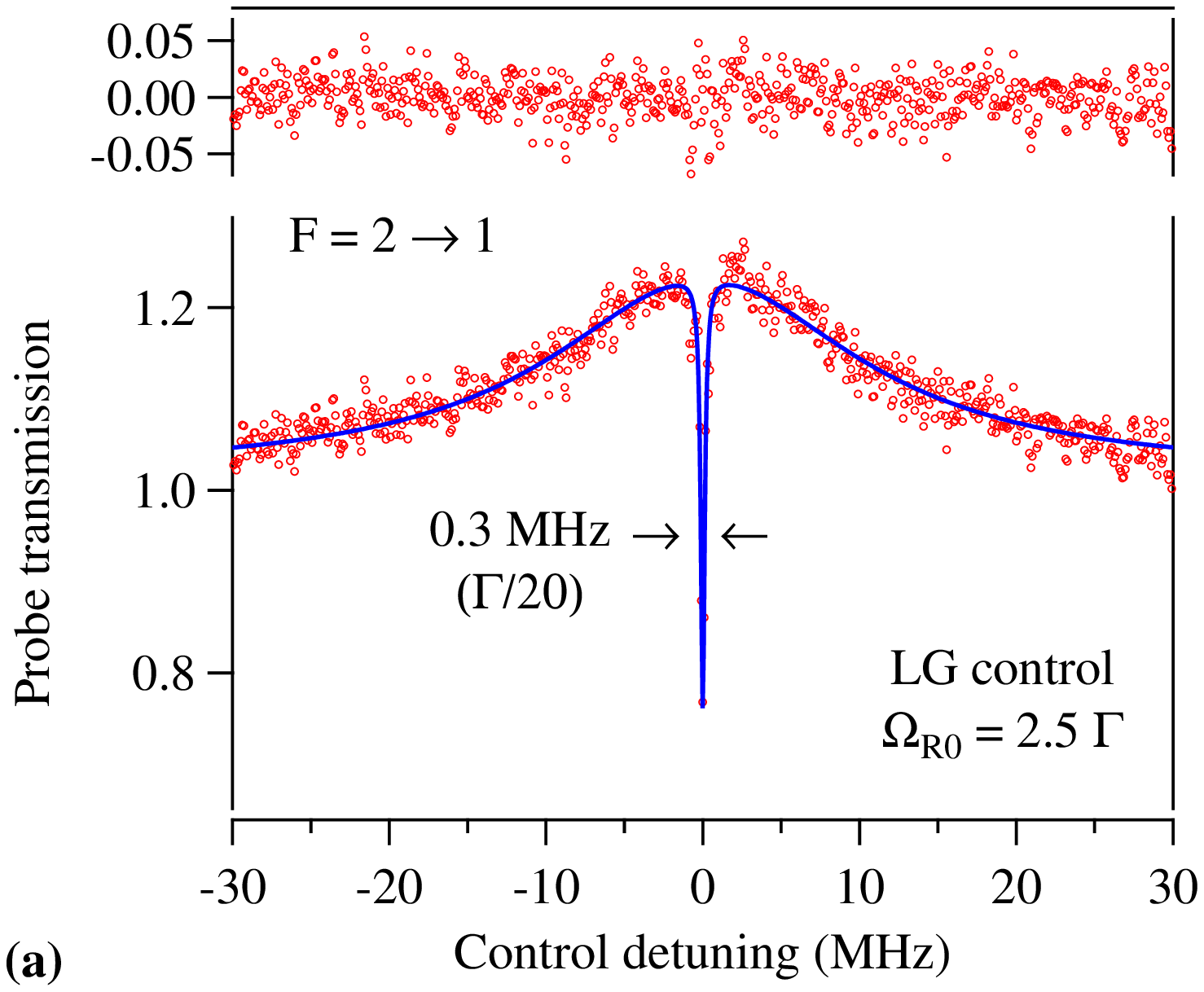}}}
\centerline{\resizebox{0.95\columnwidth}{!}{\includegraphics{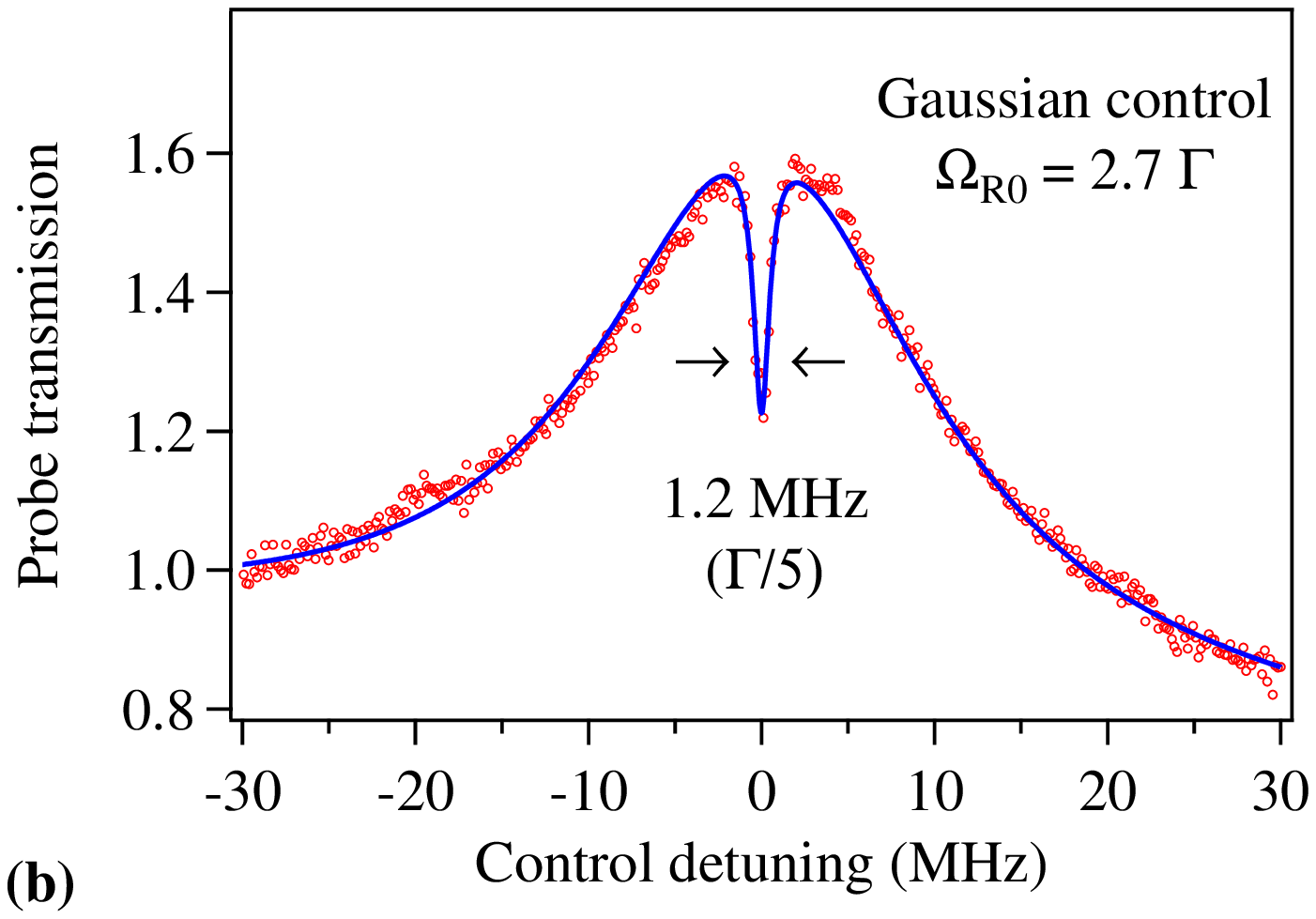}}}
\caption{(Color online) Narrow EIA resonance in the $F=2 \rightarrow 1$ transition. (a) Spectrum obtained with an LG control beam. The solid line is a curve fit with two Lorentzian lineshapes, the broad one to account for population loss and the narrow one due to EIA. The fit residuals are shown on top. (b) Similar spectrum obtained with a Gaussian control, showing a 4 times larger linewidth for the EIA dip. The maximum control Rabi frequency $\Omega_{R0}$ in each case is listed.}
 \label{eialg}
\end{figure}

The first experiments were done on the upper-level transitions. A typical probe-transmission spectrum obtained on the $F=2 \rightarrow 1$ transition with the LG control is shown in Fig.\ \ref{eialg}(a). The spectrum is normalized to make the signal equal to 1 away from line center, corresponding to absorption by zero-velocity atoms. As we approach the center, the control laser both depletes the population and saturates the transition, resulting in enhanced transmission similar to the peaks seen in saturated-absorption spectroscopy. Exactly at line center, there is a narrow enhanced absorption dip. The solid line in the figure is a curve fit to two Lorentzian lineshapes, a broad EIT peak to account for the population-depletion and saturation effects, and a narrow EIA dip which is the control-induced resonance. The featureless residuals show that the two Lorentzians describe the spectrum well, and that the noise level is less than 5\% of the signal. The full width at half maximum (FWHM) of the EIT peak is 23.8~MHz, while that of the EIA dip is only 0.3~MHz or {\em an unprecedented factor of 20 below the natural linewidth}. The maximum Rabi frequency of the control beam, corresponding to $E_0$ in Eq.\ \ref{lg}, is given by $\Omega_{R0} = 2.5 \, \Gamma$. Considering that the rms linewidth of our free-running laser is of order 1 MHz \cite{BRW01}, the fact that we can see such a narrow feature means that the linewidth of the laser when it is locked is considerably smaller. It also appears that the width of the EIA resonance can be reduced further if the linewidth of the laser can be reduced, using well-established techniques such as by locking to a high-finesse cavity.

The comparison to a spectrum obtained under similar conditions but with a Gaussian control beam is shown in Fig.\ \ref{eialg}(b). The conditions are ``similar'' in the sense that the probe power is the same, and the total control power is increased so that the maximum control Rabi frequency at the center of the Gaussian is roughly the same at $2.7 \, \Gamma$ \cite{foot1}. The general structure is the same as before, with a broad EIT peak due to population loss and a narrow EIA dip at line center. However, the EIA dip of 1.2 MHz is about 4 times wider than seen with the LG control and less deep. The broad optical-pumping EIT peak is also more prominent because of the higher control power in the region of the probe. These observations are similar to that in our earlier work in Ref.\ \cite{CSB11}.

We must contrast the narrow EIA feature seen by us with enhanced-absorption features seen in the earlier work of Ref.\ \cite{LBA99}, where also the magnetic sublevels of a transition in the $D_2$ line of Rb were used. The earlier work is like the phenomenon of coherent population trapping (CPT) in a $\Lambda$ system \cite{ARI96}, with two phase-coherent beams driving the atoms into a bright superposition state. The narrow feature (with width of about $0.1$~MHz) is seen when the difference frequency between the two beams is equal to the splitting of the ground levels, so that the ground levels are resonantly coupled to the excited level (the two-photon Raman resonance condition). In that work \cite{LBA99}, the Raman resonance condition corresponded to zero difference frequency between the two beams (because of the use of degenerate sublevels), and the phase coherence was achieved by using a single laser to generate both the beams. CPT experiments gain from the use of buffer-gas filled cells since that increases the ground-state coherence times and reduces the linewidth of the observed resonance. By contrast, the use of such cells actually kills the EIT signal \cite{foot2}. Linewidths as narrow as 42 Hz have been observed for coherent dark resonances in a Cs vapor cell filled with N$_2$ buffer gas \cite{BNW97}, but that cannot be called subnatural because the natural linewidth of the upper state of the $\Lambda$ system (5~MHz) does not enter the picture. Rather, the relevant natural width is the width of transitions between the two lower levels, which is typically less than a Hz because it is an electric-dipole-forbidden transition. Since the frequency or linewidth of the probe laser is not directly relevant, the narrow feature in CPT can only be used to stabilize {\em the phase difference} between the beams (which is the well-known clock transition in the above case). On the other hand, our {\em subnatural} resonance appears when the frequency of a laser on an optical transition is scanned. It can be used for high-resolution spectroscopy of the excited state \cite{RAN02}, tighter locking to an atomic transition, or any application where the properties (absorptive and dispersive) of an independent laser are important.

\begin{figure}
\centerline{\resizebox{0.95\columnwidth}{!}{\includegraphics{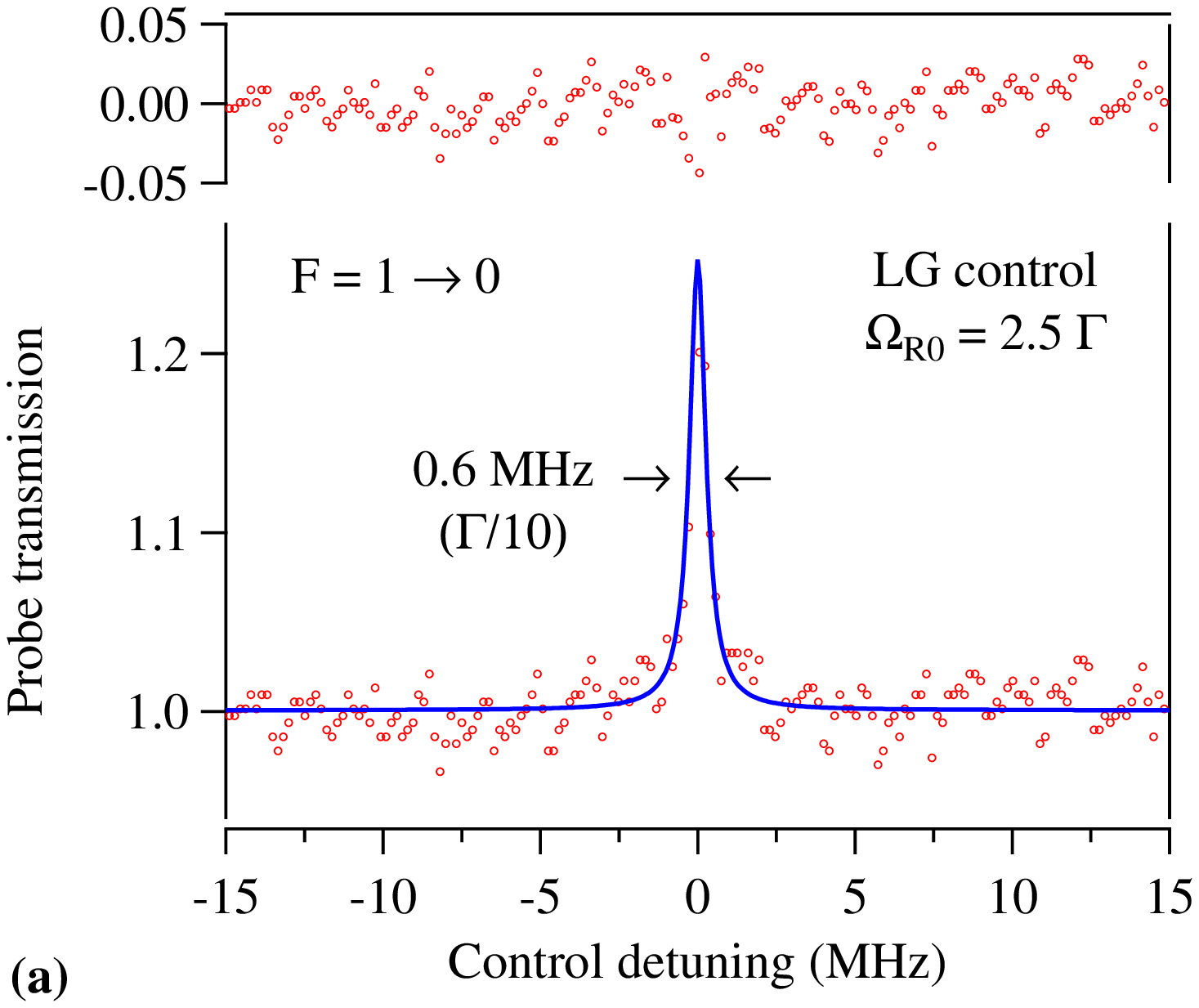}}}
\centerline{\resizebox{0.95\columnwidth}{!}{\includegraphics{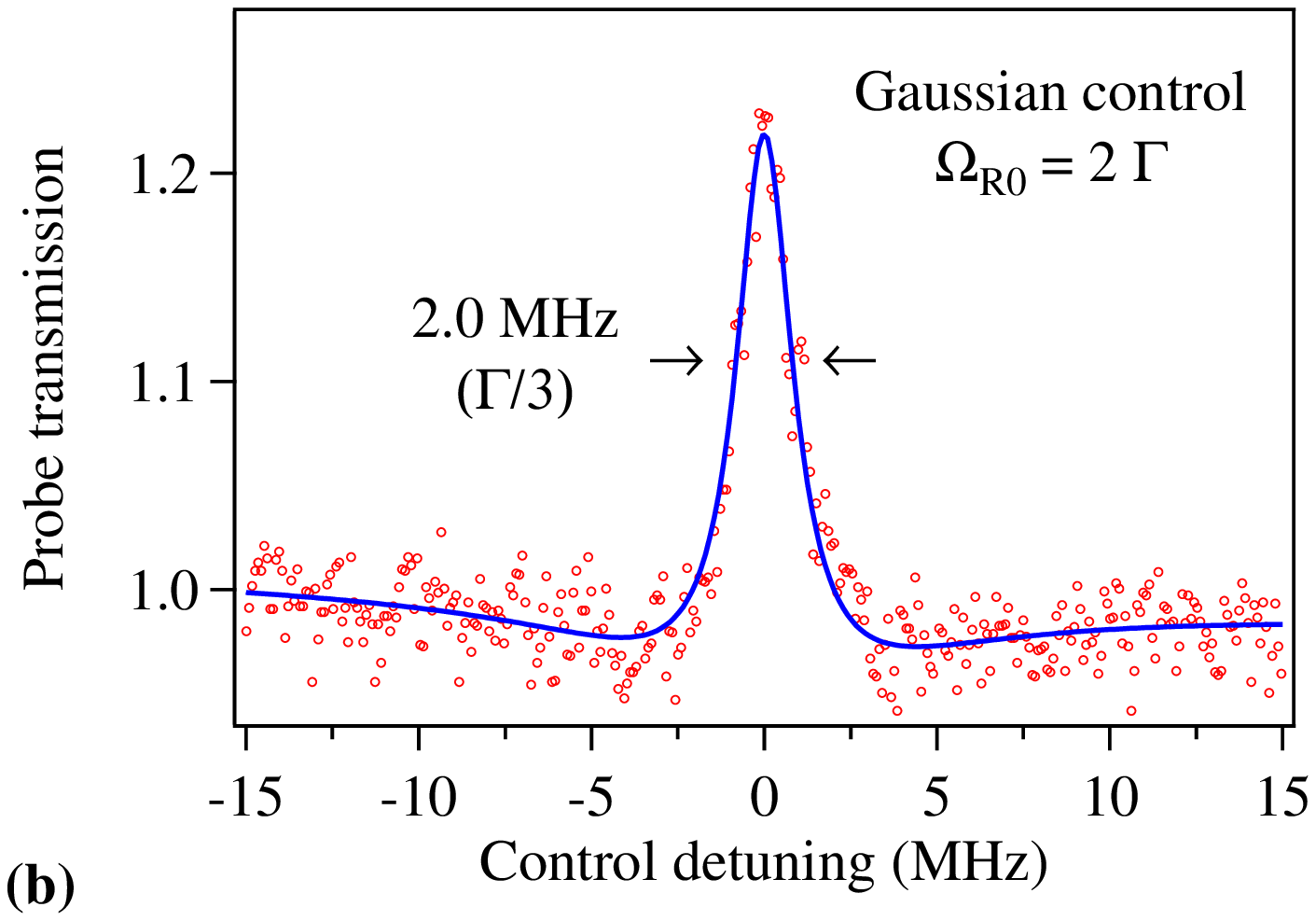}}}
\caption{(Color online) Narrowing of width of the EIT resonance in the $F=1 \rightarrow 0$ transition obtained with an LG control [shown in (a)] compared to a Gaussian control [shown in (b)]. The solid lines are curve fits to two Lorentzian lineshapes, a broad dip due to optical pumping and a narrow peak due to EIT.}
 \label{eitlg}
\end{figure}

We next turn to spectra obtained for lower-level transitions. The comparison of spectra for the $F=1 \rightarrow 0$ transition obtained with the LG control and the Gaussian control is shown in Fig.\ \ref{eitlg}. The spectra are again normalized to unit absorption away from line center, but the Gaussian spectrum is obtained with a slightly smaller value for the maximum Rabi frequency of $2 \, \Gamma$. In this case, optical pumping by the control laser near line center {\em increases} the population in the $m_F=\pm 1$ sublevels and hence causes a decrease in probe transmission, as explained in our earlier work \cite{CSB11}. This broad optical-pumping dip is more prominent for the Gaussian control and almost negligible for the LG control, which is to be expected because the Gaussian control is stronger in the region of the probe. At line center, the control-laser induced resonance now appears as an EIT peak. The FWHM is $\Gamma/3$ for the Gaussian control and a significantly smaller value of $\Gamma/10$ for the LG control. Thus the improvement in going from Gaussian to LG is almost the same as that of the upper-level transitions, but the overall widths are about $2\times$ higher. The curve fit to two Lorentzians describes the data well with featureless residuals and high signal-to-noise ratio.

The features that we observe are robust in terms of polarization and power. Similar narrow resonances are obtained if the polarization of the two beams is changed from linear to circular, i.e.\ ${\rm lin}$\,$\perp$\,${\rm lin}$ to $\sigma^+ \sigma^-$ (implemented by putting $\lambda/4$ waveplates on either side of the cell). With increase in control power, there is a slight increase in linewidth, which is similar to the behavior with a Gaussian control beam \cite{IKN08}. But over more than a factor of 3 increase in power, the resonances obtained with the LG control remain consistently below 1~MHz in width, which is something we have never seen with a Gaussian control beam \cite{CSB11}.

One possible explanation for this is that the LG beam carries orbital angular momentum (OAM), and hence provides a torque on the atomic trajectory within the beam. This increases the transit time and can result in a narrowing of the resonance if the transit time limits the linewidth. To understand the role of transit time in more detail, we have studied the EITA phenomena with an expanded Gaussian control beam, which should increase the transit time directly. The expanded Gaussian beam is 2.5 times larger, and has the same power as before. This means that the transit time is increased by this factor, while the Rabi frequency is reduced by the same factor. Both of these effects should cause a decrease in linewidth, instead we observe a {\em 20\% increase}. And this increase is there for both EIA resonances (upper-level transitions) and EIT resonances (lower-level transitions). This shows that transit time does not limit the observed linewidth.

\begin{figure}
\centerline{\resizebox{0.95\columnwidth}{!}{\includegraphics{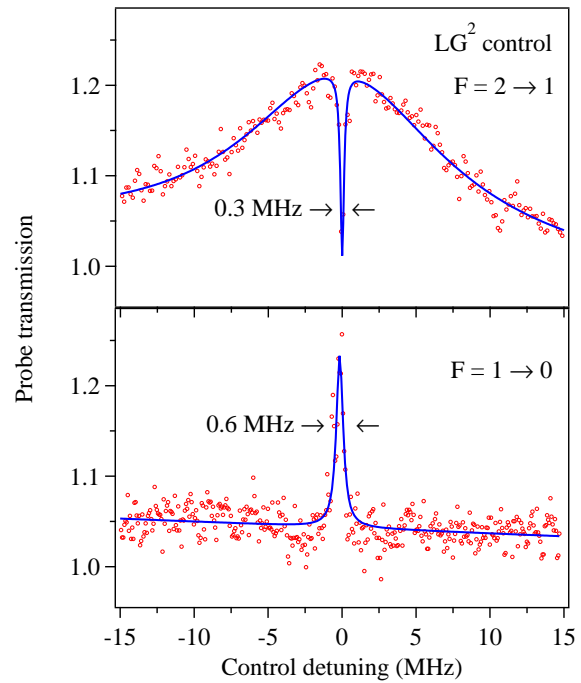}}}
\caption{(Color online) Spectra obtained with a second-order LG control beam --- EIA for the $F=2 \rightarrow 1$ transition and EIT for the $F=1 \rightarrow 0$ transition. There is no change in linewidth compared to the LG$^1$ control.}
 \label{eitalg2}
\end{figure}

We have further studied the role of OAM by using a second-order LG beam. This should double the torque on the atoms and hence increase the time spent by the atoms within the control beam. The spectra obtained with the LG$^2$ control are shown in Fig.\ \ref{eitalg2}. The important thing to note is that both the EIA and EIT features have the same linewidth as that obtained with the LG$^1$ control. One might argue that there is actually further narrowing by the LG$^2$ control but our laser linewidth is preventing us from seeing a feature narrower than 0.3 MHz (for the $F=2 \rightarrow 1$ transition). However, this is belied by the lack of narrowing of the 0.6 MHz feature for the $F=1 \rightarrow 0$ transition. We thus conclude that the OAM carried by the LG beam plays a negligible role in the linewidth reduction.

\section{Density-matrix Calculations}
In our previous work with Gaussian control beams \cite{CSB11}, we had done a complete density-matrix analysis of the sublevel structure to support the observations, especially the fact that we get EIA for transitions starting from $F=2$ and EIT for transitions starting from $F=1$. The qualitative reason for this difference was shown to be the difference in number of sublevels for the dominant transition in each set. Since this does not depend on the intensity profile of the control beam, the same difference also appears in our present work. However, there is a marked difference in the linewidth of the resonances, which is obviously related to the profile of the LG over the Gaussian. Our experiments with the expanded Gaussian also show that a weaker control beam with the same Gaussian profile will not narrow the resonance. Therefore, the particular profile of the LG mode appears important for the narrowing.

\begin{figure}
\centerline{\resizebox{0.99\columnwidth}{!}{\includegraphics{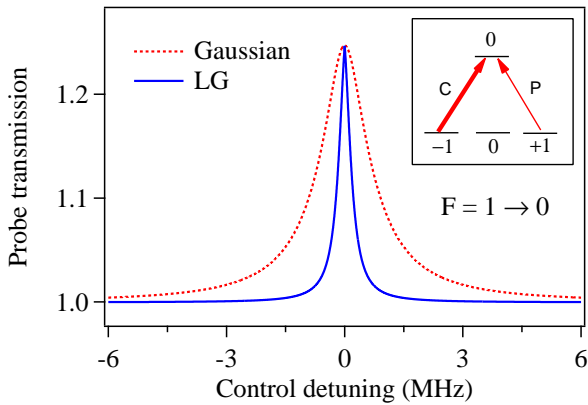}}}
\caption{(Color online) Calculated EIT spectra with LG and Gaussian control beams for the $F=1 \rightarrow 0$ transition, which forms a $\Lambda$-type system as shown in the inset.}
 \label{eitthy}
\end{figure}

To verify this explanation, we have done a density-matrix calculation with the two profiles for the $F=1 \rightarrow 0$ transition, which forms a simple $\Lambda$ system (as shown in the inset of Fig.\ \ref{eitthy}) when the beams have $\sigma^+ \sigma^-$ polarizations. We choose this because the $\Lambda$ system has been widely studied in the literature, and there is an analytic expression for probe absorption in the weak-probe regime \cite{VAR96,IKN08}. The spectra are calculated for a radially varying control Rabi frequency---varying either as Gaussian or LG---so that the radial profile leads to a radial dependence of probe absorption. The results for the two profiles are shown in Fig.\ \ref{eitthy}. In the calculation done in Ref.\ \cite{KAA00} to explain atom waveguiding by an LG beam, the spatial variation in Rabi frequency included both the amplitude and phase of the electric field. But our experiments with the LG$^2$ beam show that the phase (and resulting OAM) do not play a significant role. Therefore, the calculations are done with only the amplitude variation of the electric field. The value of $w_0$ for both profiles is taken to match the experimental value of 1.4~mm. The maximum control Rabi frequency is taken to be $0.66 \, \Gamma$. This is smaller than the experimental value because the calculation does not take into account Doppler averaging, which we have shown results in a further narrowing of the EIT resonance \cite{IKN08}.

The figure shows clearly that there is significant narrowing with the LG control, exactly as seen in our experiments. The calculation does not capture the phenomenon of optical pumping, because, in the steady state, the control laser will cause complete population transfer due to optical pumping and suppress any other feature. Neither can it capture the signal-to-noise ratio, which plays an important role in our experiment. Nevertheless, it shows that LG profile (without OAM) causes narrowing compared to the Gaussian profile.

\section{Conclusion}
In conclusion, we have studied the phenomenon of electromagnetically induced transparency and absorption using a control beam with a Laguerre-Gaussian profile instead of Gaussian, and find a marked reduction in the width of the control-induced resonances. We study these resonances using two-level hyperfine transitions on the $D_2$ line of $^{87}$Rb. The system is two level in the sense that both the control and probe lasers are on the same transition and no third level is involved, but each level has multiple magnetic sublevels. We see narrow EIA resonances for upper-level transitions and narrow EIT resonances for lower-level transitions, a difference that was explained using density-matrix analysis of the sublevel structure \cite{CSB11}. We find the same difference here, but the resonances are significantly narrower. For the $F=2 \rightarrow 1$ transition, we obtain a width that is 20 times below the natural linewidth. The physical explanation for the narrowing is the almost negligible control power for the LG beam in the region where the probe absorption is high. We show that transit time and the OAM carried by the LG beam do not play a significant role. The above explanation is verified by a density-matrix calculation for the $F=1 \rightarrow 0$ system with the two profiles. We have also studied the effect of an LG control beam for EIT in the more conventional 3-level $\Lambda$-type systems. Such systems can be formed in $^{87}$Rb using the levels $F=1 \rightarrow F'=1 \leftarrow F=2$ or the levels $F=1 \rightarrow F'=2 \leftarrow F=2$ \cite{IKN08}. We again see a reduction in linewidth compared to the use of a Gaussian beam. Similarly, the use of an LG control beam has been shown to cause a reduction in linewidth of the Hanle resonance \cite{ARP10}. Therefore, it seems that the use of an LG beam could prove advantageous in several kinds of pump-probe spectroscopy experiments.

\acknowledgments
This work was supported by the Department of Science and
Technology, Government of India. V.N. thanks G.~S.~Agarwal for useful discussions regarding calculations with LG beams.


\begin{thebibliography}{31}
\expandafter\ifx\csname natexlab\endcsname\relax\def\natexlab#1{#1}\fi
\expandafter\ifx\csname bibnamefont\endcsname\relax
  \def\bibnamefont#1{#1}\fi
\expandafter\ifx\csname bibfnamefont\endcsname\relax
  \def\bibfnamefont#1{#1}\fi
\expandafter\ifx\csname citenamefont\endcsname\relax
  \def\citenamefont#1{#1}\fi
\expandafter\ifx\csname url\endcsname\relax
  \def\url#1{\texttt{#1}}\fi
\expandafter\ifx\csname urlprefix\endcsname\relax\def\urlprefix{URL }\fi
\providecommand{\bibinfo}[2]{#2}
\providecommand{\eprint}[2][]{\url{#2}}

\bibitem[{\citenamefont{Allen et~al.}(2003)\citenamefont{Allen, Barnett, and
  Padgett}}]{ABP03}
\bibinfo{editor}{\bibfnamefont{L.}~\bibnamefont{Allen}},
  \bibinfo{editor}{\bibfnamefont{S.}~\bibnamefont{Barnett}}, \bibnamefont{and}
  \bibinfo{editor}{\bibfnamefont{M.~J.} \bibnamefont{Padgett}}, eds.,
  \emph{\bibinfo{title}{Optical Angular Momentum}} (\bibinfo{publisher}{IOP
  Publishing}, \bibinfo{address}{Bristol, UK}, \bibinfo{year}{2003}).

\bibitem[{\citenamefont{Paterson et~al.}(2001)\citenamefont{Paterson,
  MacDonald, Arlt, Sibbett, Bryant, and Dholakia}}]{PMA01}
\bibinfo{author}{\bibfnamefont{L.}~\bibnamefont{Paterson}},
  \bibinfo{author}{\bibfnamefont{M.~P.} \bibnamefont{MacDonald}},
  \bibinfo{author}{\bibfnamefont{J.}~\bibnamefont{Arlt}},
  \bibinfo{author}{\bibfnamefont{W.}~\bibnamefont{Sibbett}},
  \bibinfo{author}{\bibfnamefont{P.~E.} \bibnamefont{Bryant}},
  \bibnamefont{and} \bibinfo{author}{\bibfnamefont{K.}~\bibnamefont{Dholakia}},
  \bibinfo{journal}{Science} \textbf{\bibinfo{volume}{292}},
  \bibinfo{pages}{912} (\bibinfo{year}{2001}),
  \eprint{http://www.sciencemag.org/content/292/5518/912.full.pdf},
  \urlprefix\url{http://www.sciencemag.org/content/292/5518/912.abstract}.

\bibitem[{\citenamefont{Andersen et~al.}(2006)\citenamefont{Andersen, Ryu,
  Clad\'e, Natarajan, Vaziri, Helmerson, and Phillips}}]{ARC06}
\bibinfo{author}{\bibfnamefont{M.~F.} \bibnamefont{Andersen}},
  \bibinfo{author}{\bibfnamefont{C.}~\bibnamefont{Ryu}},
  \bibinfo{author}{\bibfnamefont{P.}~\bibnamefont{Clad\'e}},
  \bibinfo{author}{\bibfnamefont{V.}~\bibnamefont{Natarajan}},
  \bibinfo{author}{\bibfnamefont{A.}~\bibnamefont{Vaziri}},
  \bibinfo{author}{\bibfnamefont{K.}~\bibnamefont{Helmerson}},
  \bibnamefont{and} \bibinfo{author}{\bibfnamefont{W.~D.}
  \bibnamefont{Phillips}}, \bibinfo{journal}{Phys. Rev. Lett.}
  \textbf{\bibinfo{volume}{97}}, \bibinfo{pages}{170406}
  (\bibinfo{year}{2006}),
  \urlprefix\url{http://link.aps.org/doi/10.1103/PhysRevLett.97.170406}.

\bibitem[{\citenamefont{Truscott et~al.}(1999)\citenamefont{Truscott, Friese,
  Heckenberg, and Rubinsztein-Dunlop}}]{TFH99}
\bibinfo{author}{\bibfnamefont{A.~G.} \bibnamefont{Truscott}},
  \bibinfo{author}{\bibfnamefont{M.~E.~J.} \bibnamefont{Friese}},
  \bibinfo{author}{\bibfnamefont{N.~R.} \bibnamefont{Heckenberg}},
  \bibnamefont{and}
  \bibinfo{author}{\bibfnamefont{H.}~\bibnamefont{Rubinsztein-Dunlop}},
  \bibinfo{journal}{Phys. Rev. Lett.} \textbf{\bibinfo{volume}{82}},
  \bibinfo{pages}{1438} (\bibinfo{year}{1999}),
  \urlprefix\url{http://link.aps.org/doi/10.1103/PhysRevLett.82.1438}.

\bibitem[{\citenamefont{Kapoor and Agarwal}(2000)}]{KAA00}
\bibinfo{author}{\bibfnamefont{R.}~\bibnamefont{Kapoor}} \bibnamefont{and}
  \bibinfo{author}{\bibfnamefont{G.~S.} \bibnamefont{Agarwal}},
  \bibinfo{journal}{Phys. Rev. A} \textbf{\bibinfo{volume}{61}},
  \bibinfo{pages}{053818} (\bibinfo{year}{2000}),
  \urlprefix\url{http://link.aps.org/doi/10.1103/PhysRevA.61.053818}.

\bibitem[{\citenamefont{Anupriya et~al.}(2010)\citenamefont{Anupriya, Ram, and
  Pattabiraman}}]{ARP10}
\bibinfo{author}{\bibfnamefont{J.}~\bibnamefont{Anupriya}},
  \bibinfo{author}{\bibfnamefont{N.}~\bibnamefont{Ram}}, \bibnamefont{and}
  \bibinfo{author}{\bibfnamefont{M.}~\bibnamefont{Pattabiraman}},
  \bibinfo{journal}{Phys. Rev. A} \textbf{\bibinfo{volume}{81}},
  \bibinfo{pages}{043804} (\bibinfo{year}{2010}).

\bibitem[{\citenamefont{Harris}(1997{\natexlab{a}})}]{HAR97}
\bibinfo{author}{\bibfnamefont{S.~E.} \bibnamefont{Harris}},
  \bibinfo{journal}{Phys. Today} \textbf{\bibinfo{volume}{50}},
  \bibinfo{pages}{36} (\bibinfo{year}{1997}{\natexlab{a}}).

\bibitem[{\citenamefont{Fleischhauer et~al.}(2005)\citenamefont{Fleischhauer,
  Imamoglu, and Marangos}}]{FIM05}
\bibinfo{author}{\bibfnamefont{M.}~\bibnamefont{Fleischhauer}},
  \bibinfo{author}{\bibfnamefont{A.}~\bibnamefont{Imamoglu}}, \bibnamefont{and}
  \bibinfo{author}{\bibfnamefont{J.~P.} \bibnamefont{Marangos}},
  \bibinfo{journal}{Rev. Mod. Phys.} \textbf{\bibinfo{volume}{77}},
  \bibinfo{eid}{633} (pages~\bibinfo{numpages}{41}) (\bibinfo{year}{2005}),
  \urlprefix\url{http://link.aps.org/abstract/RMP/v77/p633}.

\bibitem[{\citenamefont{Chanu et~al.}(2011)\citenamefont{Chanu, Singh, Brun,
  Pandey, and Natarajan}}]{CSB11}
\bibinfo{author}{\bibfnamefont{S.~R.} \bibnamefont{Chanu}},
  \bibinfo{author}{\bibfnamefont{A.~K.} \bibnamefont{Singh}},
  \bibinfo{author}{\bibfnamefont{B.}~\bibnamefont{Brun}},
  \bibinfo{author}{\bibfnamefont{K.}~\bibnamefont{Pandey}}, \bibnamefont{and}
  \bibinfo{author}{\bibfnamefont{V.}~\bibnamefont{Natarajan}},
  \bibinfo{journal}{Opt. Commun.} \textbf{\bibinfo{volume}{284}},
  \bibinfo{pages}{4957 } (\bibinfo{year}{2011}), ISSN
  \bibinfo{issn}{0030-4018},
  \urlprefix\url{http://www.sciencedirect.com/science/article/pii/S0030401811007127}.

\bibitem[{\citenamefont{Agarwal}(1991)}]{AGA91}
\bibinfo{author}{\bibfnamefont{G.~S.} \bibnamefont{Agarwal}},
  \bibinfo{journal}{Phys. Rev. Lett.} \textbf{\bibinfo{volume}{67}},
  \bibinfo{pages}{980} (\bibinfo{year}{1991}).

\bibitem[{\citenamefont{Harris et~al.}(1990)\citenamefont{Harris, Field, and
  Imamo\ifmmode~\breve{g}\else \u{g}\fi{}lu}}]{HFI90}
\bibinfo{author}{\bibfnamefont{S.~E.} \bibnamefont{Harris}},
  \bibinfo{author}{\bibfnamefont{J.~E.} \bibnamefont{Field}}, \bibnamefont{and}
  \bibinfo{author}{\bibfnamefont{A.}~\bibnamefont{Imamo\ifmmode~\breve{g}\else
  \u{g}\fi{}lu}}, \bibinfo{journal}{Phys. Rev. Lett.}
  \textbf{\bibinfo{volume}{64}}, \bibinfo{pages}{1107} (\bibinfo{year}{1990}).

\bibitem[{\citenamefont{Hau et~al.}(1999)\citenamefont{Hau, Harris, Dutton, and
  Behroozi}}]{HHD99}
\bibinfo{author}{\bibfnamefont{L.~V.} \bibnamefont{Hau}},
  \bibinfo{author}{\bibfnamefont{S.~E.} \bibnamefont{Harris}},
  \bibinfo{author}{\bibfnamefont{Z.}~\bibnamefont{Dutton}}, \bibnamefont{and}
  \bibinfo{author}{\bibfnamefont{C.~H.} \bibnamefont{Behroozi}},
  \bibinfo{journal}{Nature (London)} \textbf{\bibinfo{volume}{397}},
  \bibinfo{pages}{594} (\bibinfo{year}{1999}).

\bibitem[{\citenamefont{Wu and Xiao}(2008)}]{WUX08}
\bibinfo{author}{\bibfnamefont{H.}~\bibnamefont{Wu}} \bibnamefont{and}
  \bibinfo{author}{\bibfnamefont{M.}~\bibnamefont{Xiao}},
  \bibinfo{journal}{Phys. Rev. A} \textbf{\bibinfo{volume}{77}},
  \bibinfo{eid}{031801} (pages~\bibinfo{numpages}{4}) (\bibinfo{year}{2008}),
  \urlprefix\url{http://link.aps.org/abstract/PRA/v77/e031801}.

\bibitem[{\citenamefont{Cui et~al.}(2007)\citenamefont{Cui, Jia, Gao, Xue,
  Wang, and Wu}}]{CJG07}
\bibinfo{author}{\bibfnamefont{C.-L.} \bibnamefont{Cui}},
  \bibinfo{author}{\bibfnamefont{J.-K.} \bibnamefont{Jia}},
  \bibinfo{author}{\bibfnamefont{J.-W.} \bibnamefont{Gao}},
  \bibinfo{author}{\bibfnamefont{Y.}~\bibnamefont{Xue}},
  \bibinfo{author}{\bibfnamefont{G.}~\bibnamefont{Wang}}, \bibnamefont{and}
  \bibinfo{author}{\bibfnamefont{J.-H.} \bibnamefont{Wu}},
  \bibinfo{journal}{Phys. Rev. A} \textbf{\bibinfo{volume}{76}},
  \bibinfo{pages}{033815} (\bibinfo{year}{2007}).

\bibitem[{\citenamefont{Niu et~al.}(2005)\citenamefont{Niu, Gong, Li, Xu, and
  Liang}}]{NGL05}
\bibinfo{author}{\bibfnamefont{Y.}~\bibnamefont{Niu}},
  \bibinfo{author}{\bibfnamefont{S.}~\bibnamefont{Gong}},
  \bibinfo{author}{\bibfnamefont{R.}~\bibnamefont{Li}},
  \bibinfo{author}{\bibfnamefont{Z.}~\bibnamefont{Xu}}, \bibnamefont{and}
  \bibinfo{author}{\bibfnamefont{X.}~\bibnamefont{Liang}},
  \bibinfo{journal}{Opt. Lett.} \textbf{\bibinfo{volume}{30}},
  \bibinfo{pages}{3371} (\bibinfo{year}{2005}),
  \urlprefix\url{http://ol.osa.org/abstract.cfm?URI=ol-30-24-3371}.

\bibitem[{\citenamefont{Rapol and Natarajan}(2002)}]{RAN02}
\bibinfo{author}{\bibfnamefont{U.~D.} \bibnamefont{Rapol}} \bibnamefont{and}
  \bibinfo{author}{\bibfnamefont{V.}~\bibnamefont{Natarajan}},
  \bibinfo{journal}{Europhys. Lett.} \textbf{\bibinfo{volume}{60}},
  \bibinfo{pages}{195–} (\bibinfo{year}{2002}).

\bibitem[{\citenamefont{Das and Natarajan}(2005)}]{DAN05}
\bibinfo{author}{\bibfnamefont{D.}~\bibnamefont{Das}} \bibnamefont{and}
  \bibinfo{author}{\bibfnamefont{V.}~\bibnamefont{Natarajan}},
  \bibinfo{journal}{Europhys. Lett.} \textbf{\bibinfo{volume}{72}},
  \bibinfo{pages}{740} (\bibinfo{year}{2005}).

\bibitem[{\citenamefont{Cohen-Tannoudji and Reynaud}(1977)}]{COR77}
\bibinfo{author}{\bibfnamefont{C.}~\bibnamefont{Cohen-Tannoudji}}
  \bibnamefont{and} \bibinfo{author}{\bibfnamefont{S.}~\bibnamefont{Reynaud}},
  \bibinfo{journal}{J. Phys. B.} \textbf{\bibinfo{volume}{10}},
  \bibinfo{pages}{365} (\bibinfo{year}{1977}).

\bibitem[{\citenamefont{Li and Xiao}(1995)}]{LIX95}
\bibinfo{author}{\bibfnamefont{Y.-q.} \bibnamefont{Li}} \bibnamefont{and}
  \bibinfo{author}{\bibfnamefont{M.}~\bibnamefont{Xiao}},
  \bibinfo{journal}{Phys. Rev. A} \textbf{\bibinfo{volume}{51}},
  \bibinfo{pages}{4959} (\bibinfo{year}{1995}).

\bibitem[{\citenamefont{Iftiquar et~al.}(2008)\citenamefont{Iftiquar, Karve,
  and Natarajan}}]{IKN08}
\bibinfo{author}{\bibfnamefont{S.~M.} \bibnamefont{Iftiquar}},
  \bibinfo{author}{\bibfnamefont{G.~R.} \bibnamefont{Karve}}, \bibnamefont{and}
  \bibinfo{author}{\bibfnamefont{V.}~\bibnamefont{Natarajan}},
  \bibinfo{journal}{Phys. Rev. A} \textbf{\bibinfo{volume}{77}},
  \bibinfo{eid}{063807} (pages~\bibinfo{numpages}{5}) (\bibinfo{year}{2008}),
  \urlprefix\url{http://link.aps.org/abstract/PRA/v77/e063807}.

\bibitem[{\citenamefont{Siegman}(1986)}]{SIE86}
\bibinfo{author}{\bibfnamefont{A.~E.} \bibnamefont{Siegman}},
  \emph{\bibinfo{title}{Lasers}} (\bibinfo{publisher}{University Science
  Books}, \bibinfo{address}{Mill Valley, CA}, \bibinfo{year}{1986}).

\bibitem[{\citenamefont{Banerjee et~al.}(2001)\citenamefont{Banerjee, Rapol,
  Wasan, and Natarajan}}]{BRW01}
\bibinfo{author}{\bibfnamefont{A.}~\bibnamefont{Banerjee}},
  \bibinfo{author}{\bibfnamefont{U.~D.} \bibnamefont{Rapol}},
  \bibinfo{author}{\bibfnamefont{A.}~\bibnamefont{Wasan}}, \bibnamefont{and}
  \bibinfo{author}{\bibfnamefont{V.}~\bibnamefont{Natarajan}},
  \bibinfo{journal}{Appl. Phys. Lett.} \textbf{\bibinfo{volume}{79}},
  \bibinfo{pages}{2139} (\bibinfo{year}{2001}).

\bibitem[{\citenamefont{Bazhenov et~al.}(1990)\citenamefont{Bazhenov,
  Vasnetsov, and Soskin}}]{BVS90a}
\bibinfo{author}{\bibfnamefont{V.~Y.} \bibnamefont{Bazhenov}},
  \bibinfo{author}{\bibfnamefont{M.~V.} \bibnamefont{Vasnetsov}},
  \bibnamefont{and} \bibinfo{author}{\bibfnamefont{M.~S.}
  \bibnamefont{Soskin}}, \bibinfo{journal}{JETP Lett.}
  \textbf{\bibinfo{volume}{52}}, \bibinfo{pages}{429} (\bibinfo{year}{1990}).

\bibitem[{\citenamefont{Heckenberg et~al.}(1992)\citenamefont{Heckenberg,
  McDuff, Smith, and White}}]{HMS92}
\bibinfo{author}{\bibfnamefont{N.~R.} \bibnamefont{Heckenberg}},
  \bibinfo{author}{\bibfnamefont{R.}~\bibnamefont{McDuff}},
  \bibinfo{author}{\bibfnamefont{C.~P.} \bibnamefont{Smith}}, \bibnamefont{and}
  \bibinfo{author}{\bibfnamefont{A.~G.} \bibnamefont{White}},
  \bibinfo{journal}{Opt. Lett.} \textbf{\bibinfo{volume}{17}},
  \bibinfo{pages}{221} (\bibinfo{year}{1992}),
  \urlprefix\url{http://ol.osa.org/abstract.cfm?URI=ol-17-3-221}.

\bibitem[{\citenamefont{Chanu et~al.}(2012)\citenamefont{Chanu, Pandey, and
  Natarajan}}]{CPN12}
\bibinfo{author}{\bibfnamefont{S.~R.} \bibnamefont{Chanu}},
  \bibinfo{author}{\bibfnamefont{K.}~\bibnamefont{Pandey}}, \bibnamefont{and}
  \bibinfo{author}{\bibfnamefont{V.}~\bibnamefont{Natarajan}},
  \bibinfo{journal}{Europhys. Lett.} \textbf{\bibinfo{volume}{98}},
  \bibinfo{pages}{44009} (\bibinfo{year}{2012}),
  \urlprefix\url{http://stacks.iop.org/0295-5075/98/i=4/a=44009}.

\bibitem{foot1}
With this Rabi frequency, the dressed states created by the control laser will be $2.7 \, \Gamma$ apart. Therefore, one does not expect the EIA resonance to be subnatural, but we have shown in Ref.\ \cite{IKN08} that the resonance gets narrowed due to Doppler averaging.

\bibitem[{\citenamefont{Lezama et~al.}(1999)\citenamefont{Lezama, Barreiro, and
  Akulshin}}]{LBA99}
\bibinfo{author}{\bibfnamefont{A.}~\bibnamefont{Lezama}},
  \bibinfo{author}{\bibfnamefont{S.}~\bibnamefont{Barreiro}}, \bibnamefont{and}
  \bibinfo{author}{\bibfnamefont{A.~M.} \bibnamefont{Akulshin}},
  \bibinfo{journal}{Phys. Rev. A} \textbf{\bibinfo{volume}{59}},
  \bibinfo{pages}{4732} (\bibinfo{year}{1999}).

\bibitem[{\citenamefont{Arimondo}(1996)}]{ARI96}
\bibinfo{author}{\bibfnamefont{E.}~\bibnamefont{Arimondo}}, in
  \emph{\bibinfo{booktitle}{Progress in Optics}}, edited by
  \bibinfo{editor}{\bibfnamefont{E.}~\bibnamefont{Wolf}}
  (\bibinfo{publisher}{Elsevier Science}, \bibinfo{address}{Amsterdam},
  \bibinfo{year}{1996}), vol.~\bibinfo{volume}{35}, pp.
  \bibinfo{pages}{257--354}.

\bibitem[{\citenamefont{Natarajan}(2012)}]{foot2}
\bibinfo{author}{\bibfnamefont{V.}~\bibnamefont{Natarajan}},
  \bibinfo{journal}{to be published}  (\bibinfo{year}{2012}).

\bibitem[{\citenamefont{Brandt et~al.}(1997)\citenamefont{Brandt, Nagel,
  Wynands, and Meschede}}]{BNW97}
\bibinfo{author}{\bibfnamefont{S.}~\bibnamefont{Brandt}},
  \bibinfo{author}{\bibfnamefont{A.}~\bibnamefont{Nagel}},
  \bibinfo{author}{\bibfnamefont{R.}~\bibnamefont{Wynands}}, \bibnamefont{and}
  \bibinfo{author}{\bibfnamefont{D.}~\bibnamefont{Meschede}},
  \bibinfo{journal}{Phys. Rev. A} \textbf{\bibinfo{volume}{56}},
  \bibinfo{pages}{R1063} (\bibinfo{year}{1997}).

\bibitem[{\citenamefont{Vemuri et~al.}(1996)\citenamefont{Vemuri, Agarwal, and
  Nageswara~Rao}}]{VAR96}
\bibinfo{author}{\bibfnamefont{G.}~\bibnamefont{Vemuri}},
  \bibinfo{author}{\bibfnamefont{G.~S.} \bibnamefont{Agarwal}},
  \bibnamefont{and} \bibinfo{author}{\bibfnamefont{B.~D.}
  \bibnamefont{Nageswara~Rao}}, \bibinfo{journal}{Phys. Rev. A}
  \textbf{\bibinfo{volume}{53}}, \bibinfo{pages}{2842} (\bibinfo{year}{1996}).

\end{thebibliography}

\end{document}